\newcommand{\be}{\begin{equation}}
\newcommand{\ee}{\end{equation}}
\newcommand{\beeq}{\begin{eqnarray}}
\newcommand{\eeeq}{\end{eqnarray}}

\def\funp{{I\!\!P}}
\documentstyle[12pt,epsfig]{article}
\newlength{\dinwidth}
\newlength{\dinmargin}
\begin{document}
\thispagestyle{empty}
\begin{center}
{\hfill DTP/96/32}\\
{\hfill March 1996}
\vskip 3 cm
{\large \bf Partonic Structure of the Pomeron\footnote
{Presented at  Cracow Epiphany
Conference, 5-6 January, 1996, to be published in Acta Physica Polonica} } 
\vskip 0.5cm
{\large K. Golec--Biernat\footnote{On leave from 
H. Niewodnicza\'nski Institute of Nuclear Physics,
 Radzikowskiego 152, 31-342 Krak\'ow, Poland} }
\vskip0.5cm
{\it  Department of Physics, University of Durham, Durham,
DH1 3LE, England}
\end{center}
\vskip1cm
\begin{abstract}
The first measurement of diffractive processes in deep inelastic scattering
at the  HERA collider is analysed in terms of  a "soft" pomeron exchange model. 
The  partonic structure of the pomeron which
emerges in this picture is determined using the QCD parton model.
The important role of the gluonic component 
in the partonic interpretation of the data is particularly emphasized.
\end{abstract}
\newpage
\setcounter{page}{1}
\section{Introduction}
The recent measurements performed by   the  H1 and ZEUS
collaborations at the DESY electron-proton
collider HERA \cite{H11}-\cite{ZEUS2} have shown   
that there is a significant 
diffractive contribution in  deep inelastic scattering.
To be precise,  diffractive deep inelastic scattering
is the  process 
\be
e + p \rightarrow e^{\prime} + X + p^{\prime} 
\label{largey} 
\ee
where there is a large 
rapidity gap between the diffractively produced system $X$ and the recoil 
proton $p^{\prime}$ (or excited proton state). In principle the scattered
electron $e^{\prime}$ and the scattered proton $p^{\prime}$ 
can be identified and their
momenta  measured. In this case the cross section 
for the diffractive
process (\ref{largey}) is characterized in the leading twist approximation 
by two structure functions ${dF_2^D / dx_{I\!\!P} dt}$ 
and ${dF_L^D / dx_{I\!\!P} dt}$
in  analogy with the decomposition of the unpolarised inclusive $ep$
cross section:
\be
{d\sigma^D \over dx dQ^2d x_{I\!\!P} dt} = 
{4\pi^2\alpha^2\over xQ^4} \left [ \left (1-y+{y^2\over2} \right )
{dF_2^D(x,Q^2,x_{I\!\!P},t)\over dx_{I\!\!P} dt}-{y^2\over 2}
{dF_L^D(x,Q^2,x_{I\!\!P},t)\over dx_{I\!\!P} dt} \right]
\label{sigmad} 
\end{equation}  
{\noindent where the kinematical variables are defined as follows: }
\begin{eqnarray}
Q^2=-q^2~~~~~~~~~~~~x={Q^2\over 2pq}~~~~~~~~~~~~~~~~~~ 
y={pq\over p_ep} \nonumber  \\
t=(p-p^{\prime})^2 ~~~~~~~~\beta={Q^2\over 2q(p-p^{\prime})} 
~~~~~~~~~~x_{I\!\!P}={x\over \beta}
\label{invar}  
\end{eqnarray}
and  $q=p_e-p_e^{\prime}$, where  $p_e,p_e^{\prime},p$  and 
$p^{\prime}$ are the four momenta
 of the initial electron, final electron, 
initial proton and the recoil proton respectively (see Fig. 1). 

No accurate determination of $t$ variable has so far been possible
for the measurement of the diffractive structure function at HERA.
Therefore the structure function integrated over $t$  
\be
\label{f23}
F_2^{D(3)}(x,Q^2,x_{I\!\!P}) = \int_{-\infty}^{0} dt~
                         {dF_2^D(x,Q^2,x_{I\!\!P},t) \over dx_{I\!\!P} dt}
\ee
was evaluated from the cross section (\ref{sigmad}).
The longitudinal component was neglected
in the  original procedure of \cite{H12,ZEUS2} because of  anticipated
smallness of the longitudinal structure function and the measured 
$y$ values, $y<0.5$.

The main observation is that the structure function $F_2^{D(3)}$
factorises 
\be
\label{fact}
F_2^{D(3)}(x,Q^2,x_{I\!\!P}) = x_{I\!\!P}^{-n}~
         F(\beta, Q^2)~,
\ee
where $\beta=x/x_\funp$.
The exponent 
$n$ is found to be independent of $\beta$ and $Q^2$, and equal to
\beeq
\label{wykl}
n&=&1.19\pm0.06(stat.)\pm0.07(syst.) 
\\
n&=&1.30\pm0.08(stat.)^{+0.08}_{-0.14}(syst.)
\eeeq
for the H1 and ZEUS measurements respectively \cite{H12,ZEUS2}.

Such a universal dependence is expected in  models of diffractive
interactions (\ref{largey}) based on the concept of an exchange
of  a pomeron - an object which carries  vacuum quantum numbers 
\cite{ING1,DOLA1}.
In that case  a virtual photon, emitted
by the incoming electron,  scatters off a colourless object  (pomeron),
which carries a small fraction $x_{I\!\!P} < 0.1$ of the incoming proton's
momentum. This kinematical condition leads to a rapidity gap
between the outgoing proton $p^{\prime}$, which stays almost
intact, and the  diffractive system $X$, produced as a result of the
virtual photon-pomeron deep inelastic scattering.
The  function $F(\beta,Q^2)$ in factorization formula (\ref{fact}) is
proportional to a pomeron structure function defined in   analogy
with the proton structure function $F_2(x,Q^2)$. The $\beta$ variable
plays a role of the Bjorken variable $x$ and, if a simple
partonic interpretation of the pomeron structure function is
valid, gives the  fraction of the  pomeron's momentum carried by a partonic
constituent of the  pomeron. In a more refined approach the QCD-improved
parton model can be applied 
and the partonic content of the pomeron can
be studied just as in the proton case.

In essence this is the Ingelman-Schlein (\cite{ING1}) model.
In this paper we show that 
the diffractive deep inelastic scattering
data from HERA are described by such a model.
In this way we investigate the partonic
structure of the pomeron.
\section{The Ingelman-Schlein model}

In this approach the diffractive structure function 
${dF_2^D / dx_{I\!\!P} dt}$ is written in the factorisable form 
\be
\label{F2IS}
{dF_{2}^D \over dx_{I\!\!P} dt}(x,Q^2,x_{I\!\!P},t)= 
         f(x_{I\!\!P},t)~ F^{I\!\!P}_{2}(\beta,Q^2)~, 
\ee
where as usual $\beta=x/x_{I\!\!P}$
and $f(x_P,t)$ is the pomeron "flux factor"
which, if the diffractively recoil system is a single proton,
has the following form  \cite{DOLA1,BERGER}
\be
\label{flux}
f(x_P,t) =  N~{B^2(t) \over 16 \pi}~ x_{I\!\!P}^{1-2 \alpha(t)}~,
\ee
where $B(t)$ describes the pomeron coupling to a proton, see (\cite{GK})
for details, and $N$ is a normalisation factor. The function
\be
\label{pomtraj}
\alpha(t)=\alpha(0) + \alpha^{\prime}~t
\ee
is the Regge trajectory of the pomeron. 
The model is based on  the 
("soft") pomeron  exchange, which means that the intercept 
$\alpha(0) =1.086$
and the slope $\alpha^{\prime} = 0.25 GeV^2$. These values were obtained from
the analysis of the high energy behaviour of  total and elastic 
hadron-hadron cross sections (\cite{DLFIT}).

As we noted above
$F^{I\!\!P}_{2}(\beta,Q^2)$ is the pomeron structure 
function with the variable $\beta$ playing the role of the 
Bjorken scaling variable. The factorization 
property (\ref{F2IS}) then follows as a direct consequence 
of the single pomeron-exchange mechanism of diffraction and  the assumption 
that the pomeron is described by a single Regge pole.

The HERA measurements are found to satisfy factorisation property
(\ref{F2IS}).
Comparison of the observed values of the exponent $n$ 
with that of formula (\ref{F2IS}), after $t$ integration, 
shows that the data 
confirms dominance of the "soft" pomeron exchange, 
since the "hard" or perturbative  
pomeron with intercept $\alpha_P(0) \approx 1.5$ 
would give much steeper behaviour proportional to
$ x_\funp^{-2}$ or so.
 
In the region 
of large $Q^2$ the pomeron structure function $F_{2}^{I\!\!P}(\beta,Q^2)$  
is expected to be described in terms of the QCD parton model  which
leads to a logarithmic violation of the Bjorken scaling, implied by
perturbative QCD.
In the leading logarithmic approximation   
the pomeron structure function is related in  a conventional 
way to the quark $q_i^\funp(\beta,Q^2)$  distributions 
in a pomeron: 
\be
F_{2}^\funp(\beta,Q^2) = 2~\sum_{i=1}^{N_f} e_i^2~\beta~  q_i^\funp(\beta,Q^2)
\label{pmod}
\ee
where $e_i$ are the quark charges and $N_f$ is a number of active flavours
(note that $q_i^\funp = \bar q_i^\funp$).  
 The quark densities $q_i^\funp(\beta,Q^2)$ and the gluon
density $g^\funp(\beta,Q^2)$ of the pomeron 
evolve in $Q^2$ 
according to the  Altarelli-Parisi evolution equations. 
However, it is necessary to
provide input 
distributions at some reference scale $Q^2=Q_0^2$, 
since only the evolution is specified by  perturbative QCD. 
The parton distributions in a pomeron at the  initial scale
$Q^2_0=4 GeV^2$ were estimated in papers\cite{CK1}-\cite{GK}
using the "soft" pomeron interaction properties 
and "triple-Regge"  phenomenology. 
In the following we present the results of the analysis of ref. \cite{GK}.

First, it was assumed that the pomeron's flavour content is the same
as that of the proton. Then, the quark distributions were parametrised
in terms of a single function, the quark-singlet distribution, 
$\Sigma^\funp(\beta)= \sum_i \beta (q_i^\funp (\beta)
                    + \bar q_i^\funp (\beta)$). 
The following parametrisation was found for $Q^2_0=4 GeV^2$
\be
\Sigma^\funp(\beta)= 0.0528~\beta^{-0.08}~(1-\beta)
                   +~0.801~\beta~(1-\beta)~.
\ee
The "soft" term, proportional to $\beta^{-0.08}$, results from the 
triple-Regge limit of the diffractive process (\ref{largey}) 
which applies when  $\beta \rightarrow 0$,  
see \cite{GK} for details. The relatively small
magnitude of the coefficient is a direct consequence of the small magnitude
of the triple-pomeron coupling. The second term, which dominates 
for $\beta \sim 1$, results from an assumption that in  this region
of $\beta$ the  "soft" pomeron couples
to a quark-antiquark pair with a form factor depending on a virtuality
of the pair, see \cite{DOLA2}. 

The gluon distribution in a pomeron has a similar  form 
\be
\label{gluinp}
\beta~ g^\funp(\beta) = 0.218~ \beta^{-0.08}~(1-\beta)+
                         ~3.30~ \beta~ (1-\beta)~.
\ee
As before the "soft" term, dominating for small $\beta$, 
results from the triple Regge limit. 
The analytical form of the second term, important for large $\beta$,
 was assumed 
to be the same as for quarks and its coefficient was calculated assuming
that the longitudinal momentum sum rule holds: 
\be
\label{mom}
     \int_0^1 d\beta~\beta~\Bigl(\sum_i  ( q_i^\funp(\beta) +
           \bar q_i^\funp(\beta) ) + g^\funp(\beta) \Bigr) =1~.
\ee
The above parametrisation allows, via the evolution equations,
the diffractive structure function (\ref{F2IS}) to be determined at any
value of $Q^2$ accepted in perturbative QCD. The results are
compared to HERA data, now given in terms of a new diffractive
structure function $F_2^{D(2)}$ defined as 
\be
\label{f2xp}
{F}_2^{D(2)}(\beta,Q^2) = \int^{x_{I\!\!P_H}}_{x_{I\!\!P_L}}
  F_2^{D(3)}(\beta,Q^2,x_{I\!\!P})\,{\rm d}x_{I\!\!P}~,
\ee
where $\beta$ is fixed during the integration.
Therefore, assuming the  simple factorisation formula (\ref{F2IS}),  
we see that the function
$F_2^{D(2)}$ is proportional to the pomeron structure function $F_2^\funp$.
We find that generally the Bjorken scaling
is observed, up to a  possible mild (logarithmic) scaling violations. 
This fact
favours a partonic interpretation of the pomeron structure
function $F_2^\funp$.

In Figure 1 the solid curves show ${F}_2^{D(2)}$, obtained using 
the above parametrisation as the initial condition for $Q^2$ evolution. 
The comparison is done for H1 and ZEUS data separately,
because of different integration limits in (\ref{f2xp}) used by these
experiments, see \cite{H12,ZEUS2} for details. 
It should be  noted that 
the increase of the computed pomeron structure functions $F_2^{D(2)}$
with increasing $Q^2$, up to the value $\beta \sim 0.4$,  is related to 
the relatively large gluon distribution (\ref{gluinp}) in a pomeron 
which has a "hard" $(1-\beta)$ spectrum at $\beta \sim 1$. 
This effect is nicely confirmed by  the data, even for
$\beta = 0.65$  the measured ${F}_2^{D(2)}$ rises with $Q^2$. 
The  above behaviour of the pomeron structure function ${F}_2^{D(2)}$
is in contrast to the behaviour of the proton structure 
${F}_2(x,Q^2)$, which starts to decrease with increasing $Q^2$ already
for $x \sim 0.1$.

The presented parametrisation, however, cannot account for the  rise
of the diffractive structure function for the highest measured value of $\beta$.
The reason is simple, the relatively hard gluon spectrum is not hard enough.   
In order to understand this, let us consider the 
equation for the logarithmic slope of  ${F}_2^{\funp}$, 
resulting from the Altarelli-Parisi equations 
\be
{\frac {\partial {F}_2^{\funp} (\beta)} {\partial log(Q^2)}} = 
{\alpha_s \over {2 \pi}} 
  \int_{\beta}^1  {dz} P_{qq} (z) {F}_2^{\funp} 
  ({\frac {\beta}{z}}) + {\alpha_s \over {2 \pi}}~\delta
   \int_{\beta}^1 {dz} P_{qg} (z) {\frac {\beta}{z}} g({\frac {\beta}{z}})~,
\ee
where $\delta = \sum_{i=1}^{N_f} e_i^2$. 
The gluon term is always positive, whilst the first one could be negative,
especialy for large $\beta$ values, 
because of virtual emissions. Thus, ${F}_2^{D(2)}$ can rise
with $Q^2$ for large $\beta$  
only when the gluon distribution is large enough at  $\beta \approx 1$.

The gluon distribution  of the pomeron at large $\beta$ is the
most difficult density  to  estimate. In fact
the analytical form of the second term in (\ref{gluinp}) 
is somewhat arbitrary and  
the momentum sum rule (\ref{mom}) used to determined  its coefficient
is also a source of controversy.
The point is that the normalisation of the pomeron flux $N$ in (\ref{F2IS})
is arbitrary: any constant factor may be transferred  from 
the pomeron flux to the pomeron structure function \cite{LANDPAR}. 
Therefore, the magnitude of the momentum sum depends on  convention.

Lacking new theoretical ideas to estimate 
the  gluon distribution  of the pomeron, 
the only way to tackle the problem is to perform a QCD fit to
the diffractive structure function data.

\newpage
\section{Pomeron parton distributions from QCD fits}

In order to clarify the gluonic content of a pomeron, we perform
a global QCD fit to the diffractive data, 
published by  the H1 and ZEUS collaborations.
A first attempt along this line was done in \cite{FITPARIS}
using only H1 data. 
We take the Regge factorisable form 
\be
\label{regfac}
F_2^{D(3)}(x,Q^2,x_\funp) = x_\funp^{-n}~ F_2^\funp(\beta,Q^2)
\ee
where the partonic form (\ref{pmod}) of $F^\funp$ is assumed.

The  $x_\funp^{-n}$ factor anticipates the dominant dependence 
of the pomeron flux (\ref{flux}) on $x_\funp$ after the $t$ integration.
 The exponent
$n$ and  parameters of the parton densities at the starting scale
$Q_0^2$ are allowed to vary simultaneously in the fit.
Leading order Altarelli-Parisi evolution is used
with four massless flavours.
The  parameter $\Lambda_{QCD}$
in the running coupling constant ${\alpha_s}$  is fixed in all fits.
The momentum sum rule (\ref{mom}) is not imposed, 
therefore, the normalization of the momentum sum,  
as well as the relative quark and gluon contributions,  
are determined by the fit itself. Only statistical errors are taken into
account.

The following forms of the singlet-quark and gluon distributions 
at the initial scale $Q_0^2=4 GeV^2$ are used
\beeq
\label{initp1}
\Sigma^\funp(\beta)&=& A_1~ \beta^{A_2}~ {(1-\beta)}^{A_3}~, 
\\ 
\label{initp2}
\beta g^\funp(\beta)&=& B_1~ \beta^{B_2}~ {(1-\beta)}^{B_3}~, 
\eeeq
whereas the nonsinglet quark distribution is set to zero.

Table 1 summarises our results.
In the first fit all seven parameters are allowed to vary, whilst 
$\Lambda_{QCD}$ is fixed at $200 MeV$.
The second fit shows how the parameters change if the value of $n$ is fixed
at the value determined by ZEUS. These two fits assume that all four 
flavours are massless,
 while in the last fit the charm  flavour $c$ is massive. 
 In this case
\be
\\
F_2^\funp(\beta,Q^2) = F_2^{\funp(u,d,s)}(\beta,Q^2) 
                  +F_2^{\funp(c)}(\beta,Q^2,m_c^2)~,
\\ 
\ee
where $F_2^{\funp(u,d,s)}$ is given by QCD formula (\ref{pmod})
with three massless flavours,  and the charm contribution
$F_2^{\funp(c)}$ is driven by the gluon distribution function
\be
F_2^{\funp(c)}(\beta,Q^2,m_c^2) = e_c^2~ \frac{\alpha_s(\mu_h^2)}{ \pi}
 \int_{a\beta}^1 { d y }~\frac{\beta}{y}~  
  C_2\left(\frac{\beta}{y},~{m_c^2\over Q^2} \right)~ g^\funp(y,\mu_c^2)~ ,
\ee
where $a = 1 + 4 m_c^2/Q^2$ and $\mu_c^2=4 m_c^2$. The form of the 
$C_2$ function as well as more details 
can be found in ref. \cite{GS}. The charm contribution is different
from zero above the threshold for $c \bar c$ pair production, 
$Q^2 (1-\beta)/\beta > 4 m_c^2$. Not that there is no $c$ quark distribution
in this approach.

The values of $\chi^2/{dof}$ confirms 
the observation made by  H1 and ZEUS that the present data are in
good agreement with the factorisable form 
(\ref{regfac}) of $F_2^{D(3)}$.
The fitted parameter $n$  is in  excellent agreement
with  the H1 value ($n=-1.19$), although  fixing it  
to the ZEUS value ($n=-1.30$)  does not lead to a significant
deterioration of the quality of the fit. 
Therefore, our analysis
supports the idea of the dominance of the "soft" pomeron exchange
in diffractive HERA data \cite{H12,ZEUS2}.
\setcounter{table}{0}
\begin{table} 
\begin{center}
\begin{tabular}{|c|r|r|r|r|}
\hline
Parameters &~~Fit 1~~&~~Fit 2~~&~~Fit 3~~  \\
\hline
\hline
$-n$ &~~1.18~~&~~1.30 {\it (fix)}  &~~1.19 \\
\hline
$A_1$ & 0.066~~& 0.021~~& 0.069\\
\hline
$A_2$ &  0.29~~&  0.11~~& 0.44\\
\hline
$A_3$ & 0.72~~&  0.84~~& 0.60\\
\hline
$B_1$ & 1.22~~&   0.72~~& 1.16\\
\hline
$B_2$ & 3.13~~&  2.51~~& 5.00\\
\hline
$B_3$ &  0.31~~&  0.27~~& 0.000\\
\hline
\hline
$\Lambda_{QCD} $ &~~~200~~~~&~~~200~~~~&~~~200~~~~   \\                
\hline
$\chi^2/dof $ &~~114/96~~&~~120/96~~&~~110/96 \\
\hline
\end{tabular}
\caption{\it The  results of the fits  to 
           $F_2^{D(3)}$ diffractive structure function data
           from H1 and ZEUS experiments. $\Lambda_{QCD}$ is fixed in all
             fits. The fitted  parameters are described in the text. Only
            statistical errors are taken into account.}
\end{center}
\end{table}
The results of our studies are compared with the data in Figure 1.
The dashed curves show $F_2^{D(2)}$ from the massless flavour fit (Fit 1),
whilst the dotted ones correspond to the fit with a  massive $c$ quark
(Fit 3). The dotted curves at the bottom of each plot show the massive charm
contribution  $F_2^{\funp(c)}$ to the $F_2^{D(2)}$ structure function
in  Fit 3 (the upper dotted curves).
The curves from both fits
give a reasonable description of the data, leading to a
persistent rise of $F_2^{D(2)}$ with $log (Q^2)$, even for the largest
value of $\beta$. It is worthwhile to pointing out
 that the massive charm description offers an additional mechanism for the
 growth of  $F_2^{D(2)}$, 
 especially for large $\beta$ and for values
 of $Q^2$  well above  the $c\bar c$ pair production threshold.

Our singlet-quark and gluon distributions 
are plotted in Figure 2 as  functions of $\beta$
for different values of $Q^2$ . The continous curves 
correspond to the
parametrisation presented in the previous section, the dashed and
the dotted ones result from the massles and massive fits respectively.
Their normalisation is determined by the choice $N=2/\pi$ in
the pomeron flux, following the convention of 
refs. \cite{DOLA1,DOLA2,COLLINS}.  
With this normalisation  the momentum sum rule (\ref{mom})
equals 1.7 for the parton distributions of the massles fit.
The contribution to this sum from gluons is
$85\%$ at $Q_0^2=4 GeV^2$, falling to $75\%$ at $Q_0^2=200 GeV^2$.

The form of the gluon density at the initial  
scale $Q^2_0=4\,{GeV}^2$ confirms
the conclusion that the persistence of the rise of
$F_2^{D(2)}$  with $log(Q^2)$ can only be reproduced by a large 
gluon (relative to quark) distribution at  $\beta \sim 1$. However, it is
worth repeating that the demand for such a gluon density cannot be 
demonstrated conclusively given the magnitude of the  errors on
the present data. 
In fact, within these errors, all our gluon distributions 
give a reasonable description of the HERA data.

\section{Summary}

The analyses presented here show that the HERA diffractive deep
inelastic scattering data can be described by a model
in which the hard scattering of  the virtual probe occurs
on a colorless object with  partonic structure. This object,
a "soft" pomeron, appears naturally in the factorisable model of
Ingelman and Schlein. The partonic distributions in the pomeron
can be determined using the QCD parton model, where the
Altarelli-Parisi evolution equations play an important role,
especially for the determination of the gluonic content of a pomeron.
Our analyses suggest that, within the present accuracy,
a broad range  of gluon distribution parametrisations is allowed.
There is  a strong indication, however, that the gluons constitute a large
component of a pomeron, carrying most of a pomeron's momentum.
It will be
interesting to see whether more precise measurements confirm this
indication. 

The potentially large gluon distribution can manifest itself in a significant
heavy $c$ flavour
contribution to the diffractive structure functions 
and may also give rise to a large longitudinal structure function
${dF_L^D / dx_\funp dt}$, which was neglected in the first analyses.

More accurate measurements of the diffractive processes in deep inelastic
scattering, over as wide a kinematical range as possible, will allow 
many important theoretically questions to be addressed. 
The region of high $\beta$
will clarify the gluon distribution problem. Measurements at very high
$y$ will allow the  study of the longitudinal component 
of the diffractive cross
section (\ref{invar}), whereas measurements at $x_\funp > 0.05$ will
test the pomeron factorisation picture, which may be violated due to
lower lying  reggeon exchanges $(f, \rho, \pi,...)$. The universality of
the extracted pomeron parton distributions may also be tested, by using
them to describe exclusive  hard diffractive scattering processes
\cite{BERSOP,DELDUCA,CFS}.

In conclusion, deep inelastic diffraction is a
promising subject for further 
experimental and theoretical studies 
in the coming years.

\section*{Acknowledgments}
A very  fruitful collaboration 
with Halina Abramowicz, Jan Kwieci\'nski and Jullian P. Phillips
on the subject presented in this paper is gratefully acknowledged.
I am par\-ti\-cu\-lar\-ly indebt to Alan D. Martin 
for a careful reading of the manuscript.
I thank the Physics Department and Grey College of the University
of Durham as well as the DESY laboratory for their warm hospitality. 
This research has been supported in part by KBN grant 
No.2P03B 231 08, Maria Sklodowska-Curie Fund II (No.PAA/NSF-94-158)
and by the EU contract number CHRX-CT92-0004/CT93-357. 


\newpage
\begin{figure}[htb]
   \vspace*{-1cm}
    \centerline{
     \psfig{figure=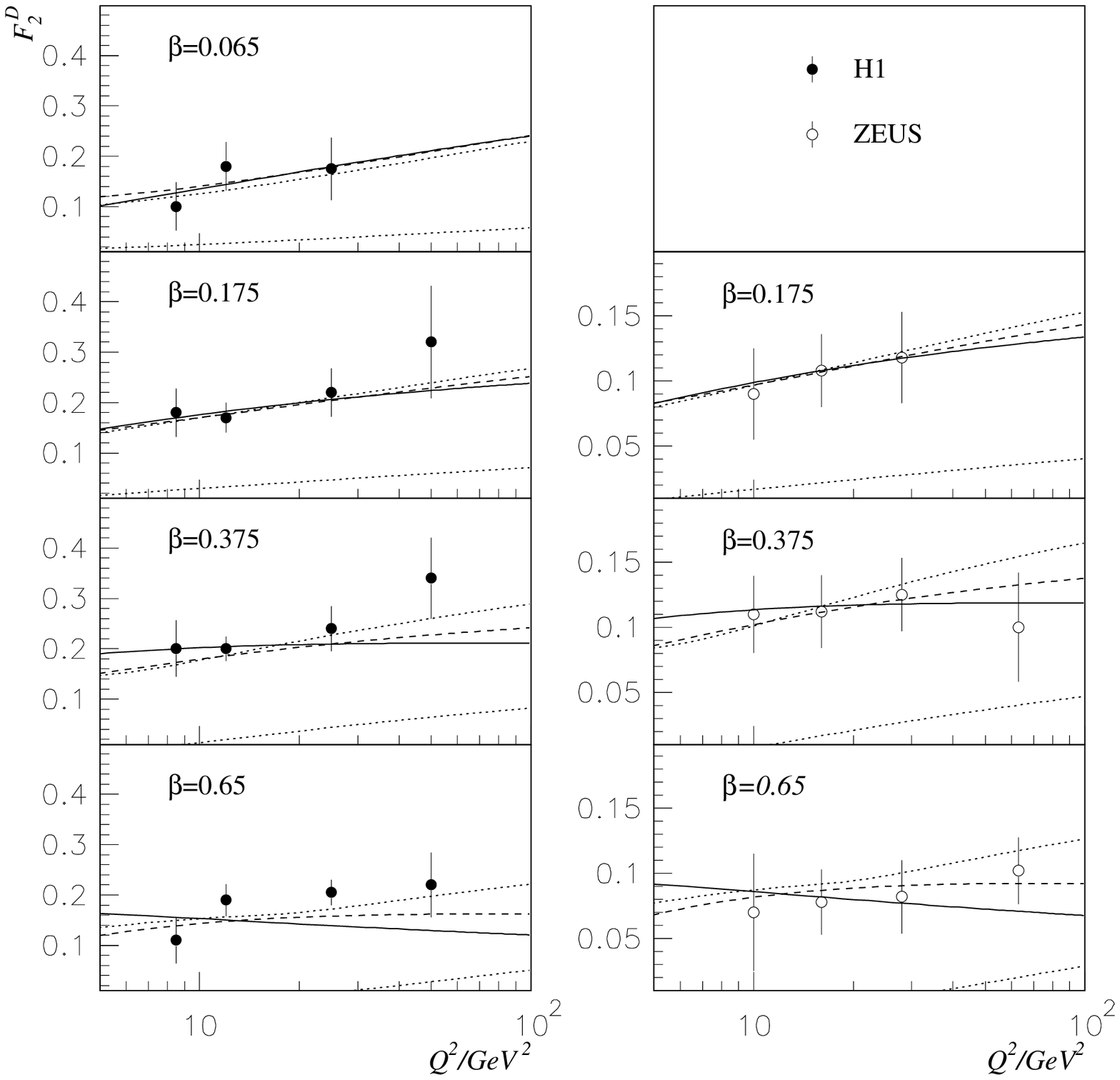,height=20cm,width=20cm}
     }
    \vspace*{-0.5cm}
     \caption{  
Comparison of the diffractive structure function $F_2^{D}$ obtained in
analyses of this paper  with H1 and ZEUS data.  
The solid curves correspond to the estimation described in
 section 2, while the dashed and dotted ones come from the massless
(Fit 1) and the massive $c$ quark (Fit 3) fits respectively.
 The dotted
curves at the bottom show the charm contribution $F_2^{D(c)}$ to the 
$F_2^{D}$ structure function obtained in the massive fit 
 (the upper dotted curves).
}
\end{figure}
\newpage
\begin{figure}[htb]
   \vspace*{-1cm}
    \centerline{
     \psfig{figure=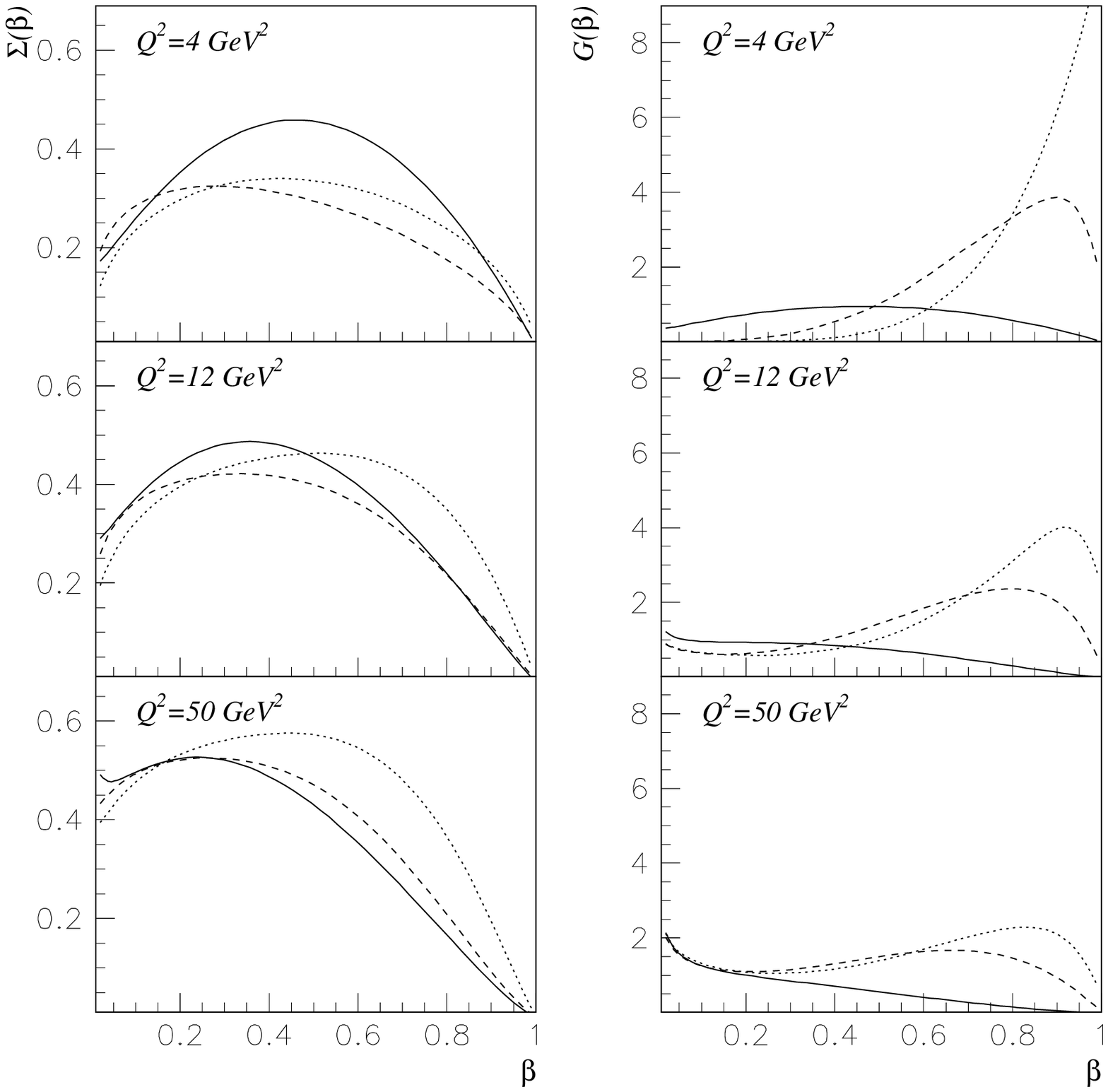,height=20cm,width=20cm}
     }
    \vspace*{-0.5cm}
     \caption{  
 The singlet and gluon distributions in a pomeron as  functions of 
 $\beta$ for different values of $Q^2$. The solid curves results from
 the estimation of chapter 2, the dashed and dotted ones are given
 by the massles (Fit 1) and massive (Fit 3) fits respectively.
 All distributions are normalised in accordance with the pomeron flux
 (\ref{flux}) normalisation $N=2/\pi$.
}
\end{figure}

\newpage
\begin{thebibliography}{999}
\bibitem{H11} H1 collaboration: T.Ahmed et al., Nucl Phys. {\bf B429} (1994) 
477.
\bibitem{H12} H1 collaboration: T. Ahmed et al., Phys. Lett. {\bf B348} (1995)
              681.
\bibitem{ZEUS1} ZEUS collaboration: M.Derrick et al., Phys. Lett. {\bf B315} 
(1993) 481; {\bf B332} (1994) 228; {\bf B338} (1994) 483.
\bibitem{ZEUS2}  ZEUS collaboration: M.Derrick et al. 
                Z.Phys. {\bf C68} (1995) 569. 
\bibitem{ING1} G. Ingelman and P.Schlein, Phys. Lett. {\bf B152} (1985) 256.
\bibitem{DOLA1} A.Donnachie and P.V.Landshoff, Nucl. Phys. {\bf B244} 
(1984)  322; {\bf B267} (1986) 690.    
\bibitem{DOLA2} A.Donnachie and P.V.Landshoff, Phys. Lett. {\bf B191} 
(1987) 309; {\bf B198} (1987) 590 (Erratum).
\bibitem{BERGER}E.L. Berger et al.,  Nucl. Phys. {\bf B286} (1987) 704.
\bibitem{COLLINS} J.C. Collins et al., Phys.Rev. {\bf D51} (1995) 3182. 
\bibitem{DLFIT} A.Donnachie and P.V.Landshoff,Phys. Lett.{\bf B296} (1992) 227. 
\bibitem{CK1} A.Capella et al., Phys. Lett. {\bf B343} (1995) 403,
              preprint LPTHE-ORSAY-95-33, hep-ph/9506454.
\bibitem{GS} T.Gehrmann and W.J.Stirling, Z.Phys. {\bf C70} (1996) 89.
\bibitem{GK} K.Golec-Biernat and J.Kwieci\'nski,
             Phys. Lett. {\bf B353} (1995) 329.
\bibitem{LANDPAR} P.V.Landshoff, in Proceedings of the Workshop DIS and QCD,
Paris 1995, editors J.-F.Laporte and Y.Soris, \'Ecole Polytechnique Edition.
\bibitem{FITPARIS} J.Dainton, J.Phillips, in Proceedings of the Workshop DIS 
and QCD,Paris 1995, editors J.-F.Laporte and Y.Soris, \'Ecole Polytechnique
Edition.
\bibitem{BERSOP}  A.Berera, D.E.Soper, preprint PSU-TH-163, Sept. 1995, 
                  hep-ph/9509239.
\bibitem{DELDUCA} V. Del Duca, talk at $10^{th}$ Topical Conference on
                  Proton-Antiproton
                  Collider Physics, Batavia, IL, 9-13 May 1995, hep-ph/9506355. 
\bibitem{CFS} J.C.Collins, L.Frankfurt and M.Strikman, Phys.Lett.
              {\bf B307} (1993) 161.


\end{thebibliography}
\end{document}